\documentclass[twocolumn]{aastex62}

\newcommand{\gaia}{\texttt{Gaia}}

\submitjournal{ApJS}

\shorttitle{M type stars catalog}
\shortauthors{Zhong \& Li et al.}

\begin{document}
\title{ Value-added catalogs of M type stars in LAMOST DR5}

\correspondingauthor{Jing Li}
\email{lijing@shao.ac.cn, jzhong@shao.ac.cn}

\author{Jing Zhong }
\affiliation{Key Laboratory for Research in Galaxies and Cosmology,
Shanghai Astronomical Observatory,
Chinese Academy of  Sciences, 80 Nandan Road, Shanghai 200030, China}

\author[0000-0002-4953-1545]{Jing Li }
\affiliation{Physics and Space Science College,China West Normal University,
1 ShiDa Road, Nanchong 637002  , China}
\affiliation{Key Laboratory for Research in Galaxies and Cosmology,
Shanghai Astronomical Observatory,
Chinese Academy of  Sciences, 80 Nandan Road, Shanghai 200030, China}
\affiliation{Chinese Academy of Sciences South America Center for Astronomy, National Astronomical Observatories, CAS, Beijing 100012, China}

\author{Jeffrey L. Carlin}
\affiliation{LSST, 950 North Cherry Avenue, Tucson, AZ 85719, USA}

\author{Li Chen }
\affiliation{Key Laboratory for Research in Galaxies and Cosmology,
Shanghai Astronomical Observatory,
Chinese Academy of  Sciences, 80 Nandan Road, Shanghai 200030, China}
\affiliation{School of Astronomy and Space Science, University of Chinese Academy of Sciences, No. 19A, Yuquan Road, Beijing 100049, China}

\author{Rene A. Mendez}
\affiliation{Departamento de Astronomia, Universidad de Chile, Casilla 36-D, Correo Central, Santiago, Chile}

\author{Jinliang Hou }
\affiliation{Key Laboratory for Research in Galaxies and Cosmology,
Shanghai Astronomical Observatory,
Chinese Academy of  Sciences, 80 Nandan Road, Shanghai 200030, China}
\affiliation{School of Astronomy and Space Science, University of Chinese Academy of Sciences, No. 19A, Yuquan Road, Beijing 100049, China}


\begin{abstract}
In this work, we present new catalogs of M giant and M dwarf stars from the LAMOST DR5. In total, 39,796 M giants and 501,152 M dwarfs are identified from the classification pipeline. The template-fitting results contain M giants with 7 temperature subtypes from M0 to M6, M dwarfs with 18 temperature subtypes from K7.0 to M8.5 and 12 metallicity subclasses from dMr to usdMp. We cross-matched our M-type catalog with the 2MASS and $WISE$ catalog to obtain infrared magnitude and colors. Adopting the distances derived from the parallaxes in \gaia{} DR2, the $M_{G}$ vs. $(G_{bp}-G_{rp})_0$ diagram shows that there are also early-type stars and white dwarf-M dwarf binaries included in our M type stars sample, with a contamination rate of about $4.6\%$ for M giants and $0.48\%$ for M dwarfs. We found that CaH spectral indices are an efficient selection criteria for carbon stars. A total of 289 carbon stars were identified from the M giants sample, and further confirmed by LAMOST spectra.

\end{abstract}



\section{Introduction}
\label{sec:intro}
M giant stars are evolved objects with high luminosity (log $L/L_{\sun}$ $\sim$ 3-4), which enables them to be detected at large distances. Therefore, M giant stars are suitable as tracers for discovering and identifying the remnants of stellar streams in the Galactic halo, revealing the accretion and merging history of the Milky Way.

In the past two decades, a set of large-scale survey projects \citep[e.g. 2MASS, UKIDSS, WISE, Pan-STARRS, \gaia][]{} gathered photometric data for thousands of M giants, which provide key support for the study of substructures, especially for the Sagittarius stream. By selecting candidates from the 2MASS, M giants were first used to map out the global view of the Sagittarius Dwarf Galaxy \citep{2003ApJ...599.1082M}. More recently, using UKIDSS, \citet{2014AJ....147...76B} established a distant M giant sample to explore the Sagittarius accretion history in the outer halo of our Galaxy.

Compared to the photometrically derived samples, there are advantages to study Galactic substructures by using spectral data of M giants. First of all, the spectra provide more features for selecting a pure sample. Secondly, an estimate of the radial velocity derived from the spectra provides an important evidence for substructure membership identification. Third, chemical abundances measured from the spectra are more reliable, and thus useful for a more detailed study of substructures. Consequently, a spectral sample of M giants with large numbers and wide sky coverage has a great scientific value for the study of Galactic substructures.

A small catalog of M giants selected using SDSS and LAMOST DR1 spectra were used to detect members of the Sagittarius stream and identify new substructures \citep{2007ApJ...670..346C,2014AJ....147...76B,2016RAA....16..125L,2019arXiv190207861L}.
In order to make optimal use of the M giant sample in LAMOST, a clean and relatively complete M giant catalog with high precision distance estimates is highly desirable.

In this work, we present a much larger catalog of spectroscopic M giants than those available in the literature. The paper is organized as follows. In section 2, we describe how we select M giants and M dwarfs from LAMOST DR5. In section 3, we describe the separation between M giants, M dwarfs, and carbon stars using photometric and spectroscopic index features. A summary is presented in the last section.

\section{Candidates selection in LAMOST DR5}

As the most efficient spectroscopic survey telescope, the LAMOST, also named as Guo Shou Jing Telescope, has finished the first stage of its regular survey (LAMOST-I; 2011-2017; including the pilot survey), and provided 9,017,844 low-resolution (R $\sim$ 1,800) optical spectra in its fifth data release, of which 8,171,443 are stellar spectra. \citep{Cui2012,Zhao2012,Luo2012}

In order to identify M-type stars from the LAMOST database, \citet{2015AJ....150...42Z} developed an automatic template-fitting algorithm to classify late-type stars. After performing the rest frame correction and the pseudo-continuum normalization of LAMOST spectrum in the range 6,000-8,000\AA, $\chi^2$ values between the target spectrum and M-dwarf template spectra were calculated, with the best fit template defined as the template spectrum which corresponds to the minimum $\chi^2$ value. Because of the low surface temperature of M-type stars, the obvious characteristics of M-type spectra are molecular absorption features (e.g., CaH, TiO, VO). Stars were rejected when they do not display any molecular absorption features and the best fit template is a non-M dwarf template (e.g., K-type stars or hotter stars). Only stars whose spectrum has molecular absorption features were labelled as M-type stars.

Using the M-type spectra identified from LAMOST DR1, \citet{2015RAA....15.1154Z} found that M giants and M dwarfs can be well discriminated in the CaH2+CaH3 vs.TiO5 spectral indices diagram. Then, a set of M giant templates with high signal-to-noise ratio were assembled, with spectral subtypes ranging from M0 to M6. Combining with the M giant templates and the previous M dwarf templates \citep{2015AJ....150...42Z}, an extended set of M-type spectral templates were used to perform the M-type stars identification and classification from LAMOST DR1. From that work, a  total of 8,639 M giants as well as 101,690 M dwarfs were positively classified by the template-fitting algorithm \citep{2015RAA....15.1154Z}.

Furthermore, using the confirmed M giants sample from LAMOST, \citet{2016ApJ...823...59L} found that the combination of WISE+2MASS bands is more efficient at selecting M giant samples than the 2MASS bands alone. Based on the revised infrared selection criteria, a relatively complete photometric sample of M giants was determined. Then, after combining with APOGEE data, a strong correlation between the W1-W2 color and [M/H] was found for M giants. Given the derived relation, \citet{2016ApJ...823...59L} estimated the photometric metallicity of M giants and further derived their photometric distances, which used the LMC, the SMC, and the Sagittarius core for calibration. The derived photometric distance uncertainties were estimated at 20\%.

In this work we have adopted the same template-fitting pipeline combined with a classification algorithm \citep{2015AJ....150...42Z} and revised M-type spectral templates \citep{2015RAA....15.1154Z} to identify and classify M-type stars from LAMOST DR5. Although the optimized template-fitting algorithm largely eliminates the majority of noisy spectra in LAMOST, a small fraction of weird spectra still remain in the M-type stars sample; most of these suffer low S/N or serious sky line contamination. To further purify the M-type stars sample, candidates have to meet the following two additional criteria: (i) the mean S/N in the range 6,000-8,000 \AA~ must be greater than 5; (ii) the spectral indices must be located on the M-type stars locus (0 $<$ TiO5 $<$ 1.2 $\&$ 0.6 $<$ CaH2+CaH3 $<$ 2.4). After excluding outlier spectra and combining duplicated spectra, a total of 540,948 spectra were identified as M-type stars, including 39,796 M giant spectra and 501,152 M dwarf spectra.

\begin{figure*}
   \centering
   \includegraphics[angle=0,scale=0.4]{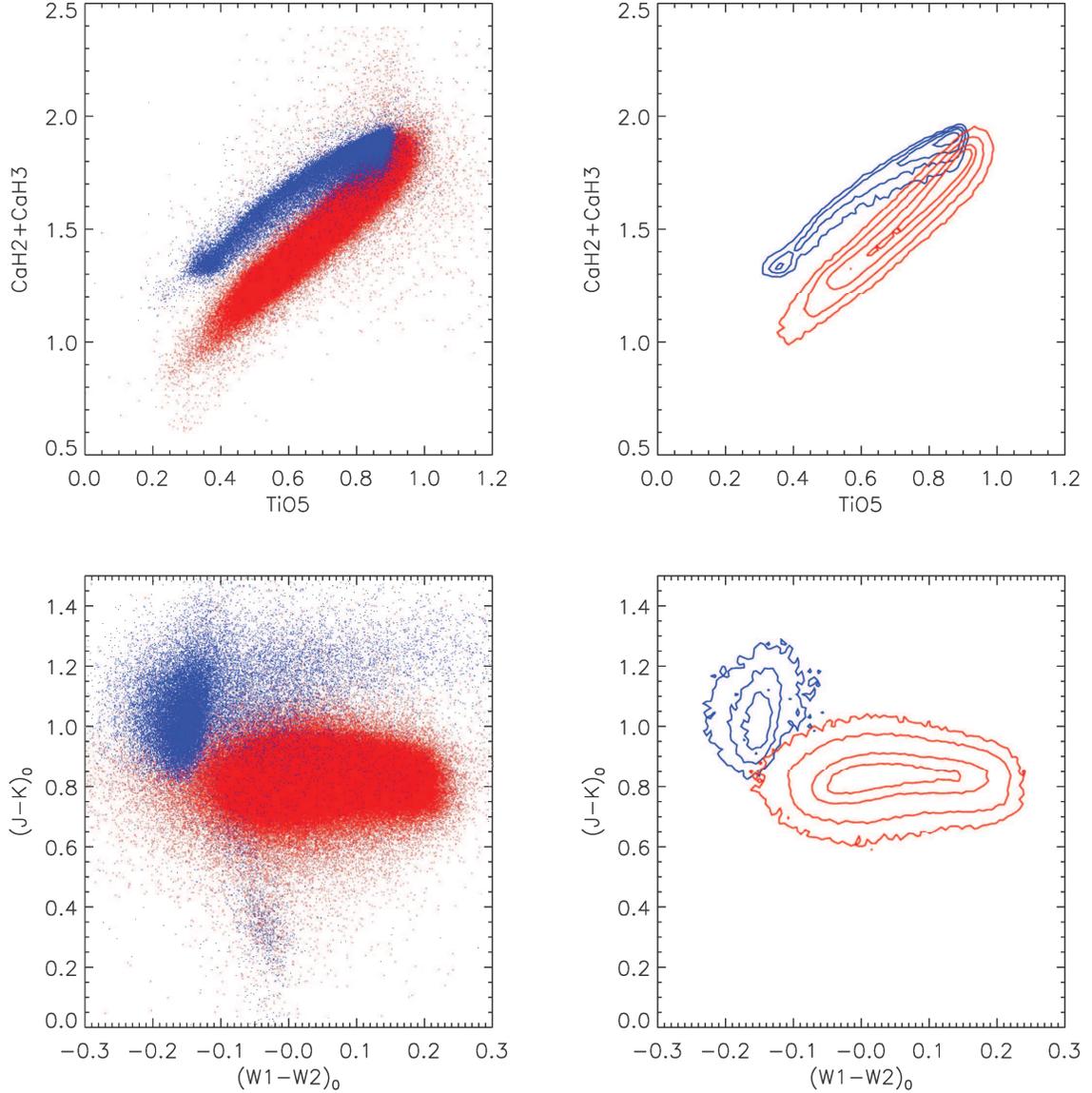}
   \caption{ Spectral indices diagram (upper panels) and de-reddened color-color diagram (bottom panels) of all M-type stars. Point distributions are plotted on the left while corresponding contours are plotted on the right. Blue and red colors represent M giants and M dwarfs respectively. As expected, M giants and M dwarfs show different distributions both in the spectroscopic parameter space as well as in the photometric parameter space.
   }
   \label{contour}
\end{figure*}

\begin{figure*}
   \centering
   \includegraphics[angle=0,scale=0.4]{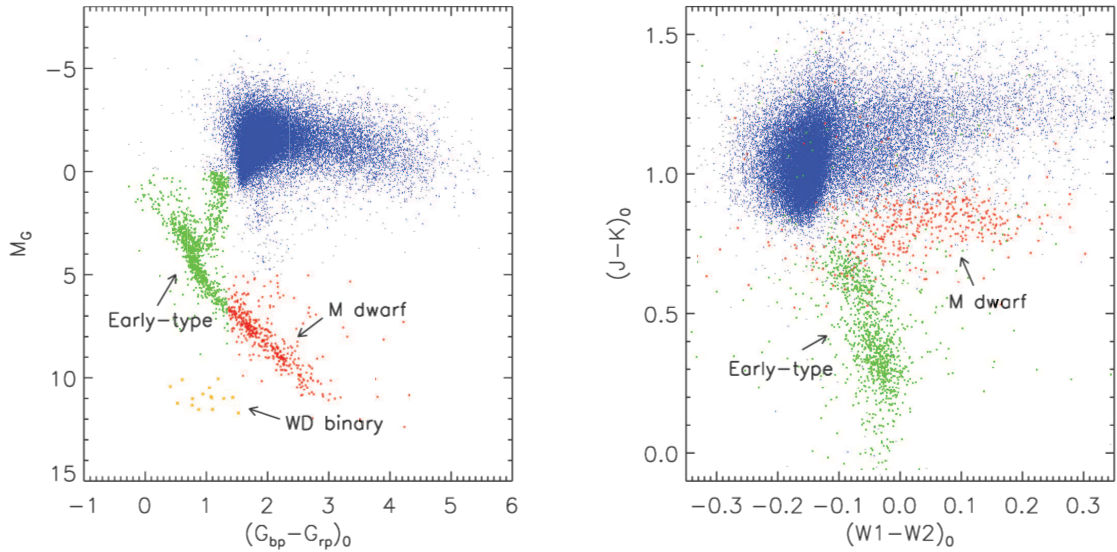}
   \caption{Color-magnitude distribution of all stars identified as M giants. Left: G-band absolute magnitude as a function of $(G_{bp}-G_{rp})_0$ color. Right: De-reddened $(J-K)_0$ vs. $(W1-W2)_0$ color-color diagram. Note that {\it all} dots are M giant candidates that were classified through LAMOST spectra. We use colors to distinguish misidentified stars, which include early-type stars (green), M dwarfs (red) and white dwarf binaries (orange).}
   \label{cmd_mg}
\end{figure*}

\begin{figure*}
   \centering
   \includegraphics[angle=0,scale=0.4]{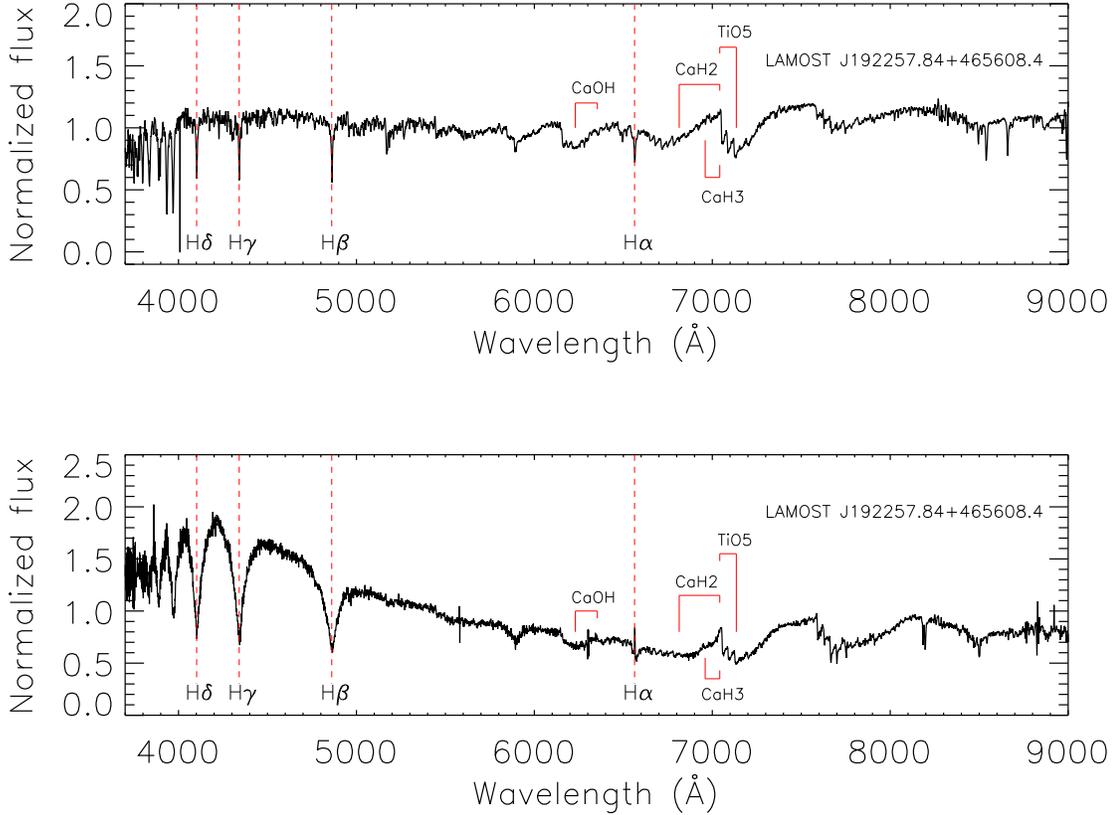}
   \caption{Two example composite spectra observed by LAMOST. Upper panel: main sequence-M giant binary. The molecular bands suggest a classical M giant spectrum in the red part, while the Balmer absorption lines suggest an F-type main sequence spectrum in the blue part.  In our M giants catalog, early-type binaries were labeled as non-M giants. Bottom panel: white dwarf-M dwarf binary, which show pressure-broadened Balmer series in the blue part and strong molecular lines in the red part. We note that these stars are labeled as white dwarf binaries in the M giants catalog. }
   \label{bexp}
\end{figure*}

\begin{figure*}
   \centering
   \includegraphics[angle=0,scale=0.4]{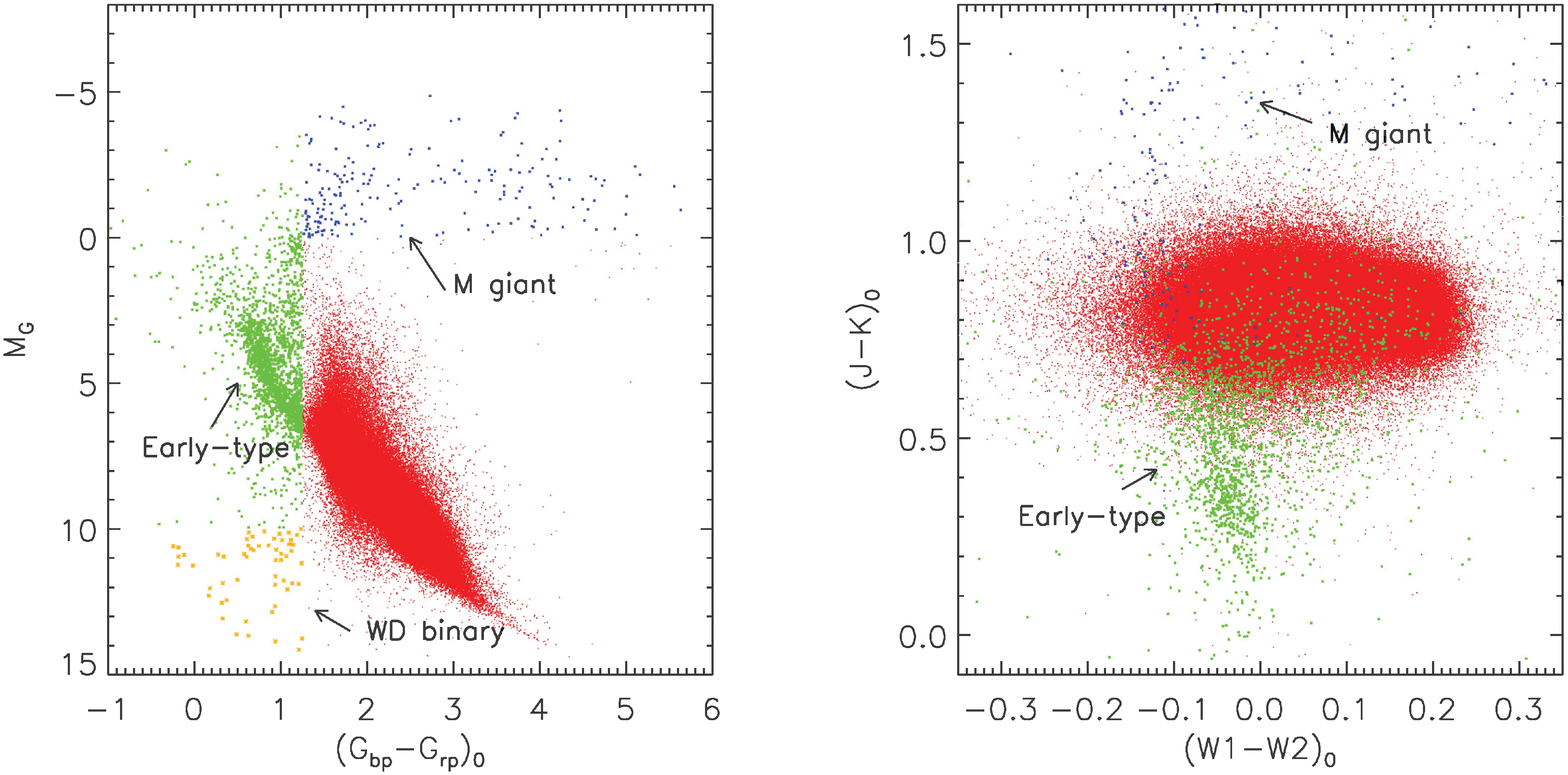}
   \caption{Color-magnitude distribution of {\it all} stars identified as M dwarfs. The panels are as described in Figure~\ref{cmd_mg}.  Misidentified stars can be separated into three groups, including early-type stars (green), M giants (blue) and white dwarf binaries (orange).}
   \label{cmd_md}
\end{figure*}

\section{Data analyses and classification results}
\label{results}
\subsection{M giants and M dwarfs}
\label{mstar}
As we described in \citet{2015RAA....15.1154Z, 2016ApJ...823...59L}, M giants and M dwarfs can be well segregated through the spectral indices diagram and the infrared color-color diagram. To verify the reliability of our classification results, we re-checked the distribution of M giants and M dwarfs with spectral indices and infrared colors as described below.

First, we cross-matched our M-type catalog with the 2MASS \citep{2006AJ....131.1163S} and \emph{WISE} \citep{2010AJ....140.1868W} infrared catalogs, using a search radius of 3$"$. In order to correct the reddening of infrared colors, a 3D Galactic dust map is needed. Since the stellar distance is an important input parameter for calculating reddening in 3D dust maps, we adopted the estimated distances from \citet{2018AJ....156...58B}, which were derived from the \gaia{} DR2 parallaxes \citep{2018A&A...616A...1G}. After further cross-matching our catalog with the value-added catalog \citep{2018AJ....156...58B}  in \gaia{} DR2 , the final common catalog contains 37,383 M giants and 472,418 M dwarfs. These number represent a loss of 6.0\% M giants and 5.7\% M dwarfs compared with the M-type star catalog in LAMOST DR5. This is because the magnitude of the excluded LAMOST M-type stars are located outside of the common magnitude range (e.g., either too faint for 2MASS/\emph{WISE} catalog or too bright for \gaia{} catalog), or because the parallax uncertainty of those excluded stars is too large to derive a reliable distance \citep{2018AJ....156...58B}.

In order to derive the intrinsic $(J-K)_0$ and (W1-W2)$_0$ of each M-type star, using the common catalog with multi-band photometry and stellar distance, we adopted the Galactic reddening E(B-V) in 3D dust maps of \citet{2018MNRAS.478..651G}, in combination with the extinction coefficients which were determined by \citet{2013MNRAS.430.2188Y}. The extinction in a given band were calculated as A$_i$=R$_i$ $\times$ E(B-V), where R$_i$ is defined as the extinction coefficient of $i$ band relative to E(B-V). In order to estimate extinction and further derive de-reddened infrared magnitude $J_0$, $H_0$, $Ks_0$, W1$_0$ and  W2$_0$, the
adopted extinction coefficient R$_i$ for each band were 0.72, 0.46, 0.306, 0.18, and 0.16, respectively \citep{2013MNRAS.430.2188Y}.

In Figure~\ref{contour} we use M-type stars in the common catalog to plot the spectral indices diagram (upper panels) and the de-reddened $(J-K)_0$ versus (W1-W2)$_0$ color-color diagram (bottom panels). Throughout panels, the distributions of M giants and M dwarfs are represented as blue and red colors respectively, both of which were identified by our template-fitting algorithm. As expected, the majority of M giants and M dwarfs are well-separated in both the spectroscopic and photometric parameter space, which also clearly illustrates the different properties of these two populations. However, there are a few blue dots located on the lower color region with $(J-K)_0$ $<$ 0.8 in the lower-left panel. As pointed out in \citet{2016ApJ...823...59L}, stars located on the blue region of the $(J-K)_0$ intrinsic color are more likely to be K-type stars, but have been  misclassified as M giants in our sample.

To further study the contaminants in our M giants sample, we calculate the absolute magnitude $M_G$ and intrinsic $(G_{bp}-G_{rp})_0$ color of each M giant star. Distance modulus and 3D dust maps were determined by \citet{2018AJ....156...58B} and \citet{2018MNRAS.478..651G}, while the extinction $A_G$ and reddening $E(BP-RP)$ were calculated as $A_G$=$2.74 \times E(B-V)$ and $E(BP-RP)=1.339 \times E(B-V)$ \citep{2018MNRAS.479L.102C}. The extra photometric and astrometric information provides a means of cross-validation of the spectral types and their properties.

Figure~\ref{cmd_mg} shows the distribution of M giants which were identified by our template-fitting algorithm. In the left panel, stellar locations in the M$_G$ versus  $(G_{bp}-G_{rp})_0$ diagram  \citep{2018A&A...616A..10G} indicate that there are indeed contaminants in the M giants sample, including early-type stars, M dwarfs, and a few white dwarfs. We replot the de-reddened $(J-K)_0$ versus (W1-W2)$_0$ color-color diagram of M giants in the right panel. Early-type stars and M dwarfs which were labeled with green and red dots in the left panel also show different distributions from bona fide M giants in the infrared color-color diagram. Comparing with the distribution of K/M stars in \citet{2016ApJ...823...59L}, the location of each stellar component is consistent with the reported region. Furthermore, we performed close visual inspection of the LAMOST spectra for each of these non-M giants. We find that a large fraction of identified early-type star candidates are main sequence-M giant binaries, and all labeled white dwarfs are white dwarf-M dwarf binaries. In Figure~\ref{bexp}, we plot LAMOST spectra of these two kinds of binaries as examples. In particular, the early-type stars and M dwarfs in the M giants sample were labeled as non-M giants and the white dwarfs were labeled as white dwarf binaries in the M giants catalog (see Section~\ref{catalog}).

The total number of non-M giant stars is 1,712 (including white dwarf binaries), corresponding to a contamination rate of about 4.6$\%$ in the M giants sample, which is similar with the contamination rate (4.7$\%$) reported in \citet{2015RAA....15.1154Z}.

Similarly, figure~\ref{cmd_md} shows the color-magnitude distribution of all identified M dwarfs. In order to exclude misidentified sources from the M dwarfs sample, we use the following criteria to separate contaminants into three groups: $(G_{bp}-G_{rp})_0 < 1.25~ mag  ~\& ~  M_G < 10 ~mag $ for early-type sources;  $(G_{bp}-G_{rp})_0 < 1.25~ mag  ~\& ~  M_G > 10 ~mag $ for white dwarf sources; $(G_{bp}-G_{rp})_0 > 1.25~ mag  ~\& ~  M_G < 0 ~mag $ for M giant sources. For those non-M dwarfs sources, we performed visual inspection of their corresponding LAMOST spectra. We find that a large number of non-M dwarf sources which were labeled as early-type or white dwarf are binaries. The total number of non-M dwarfs sources is 2281, corresponding to a contamination rate of 0.48\% in the M dwarfs sample.

\begin{figure*}
   \centering
   \includegraphics[angle=0,scale=0.45]{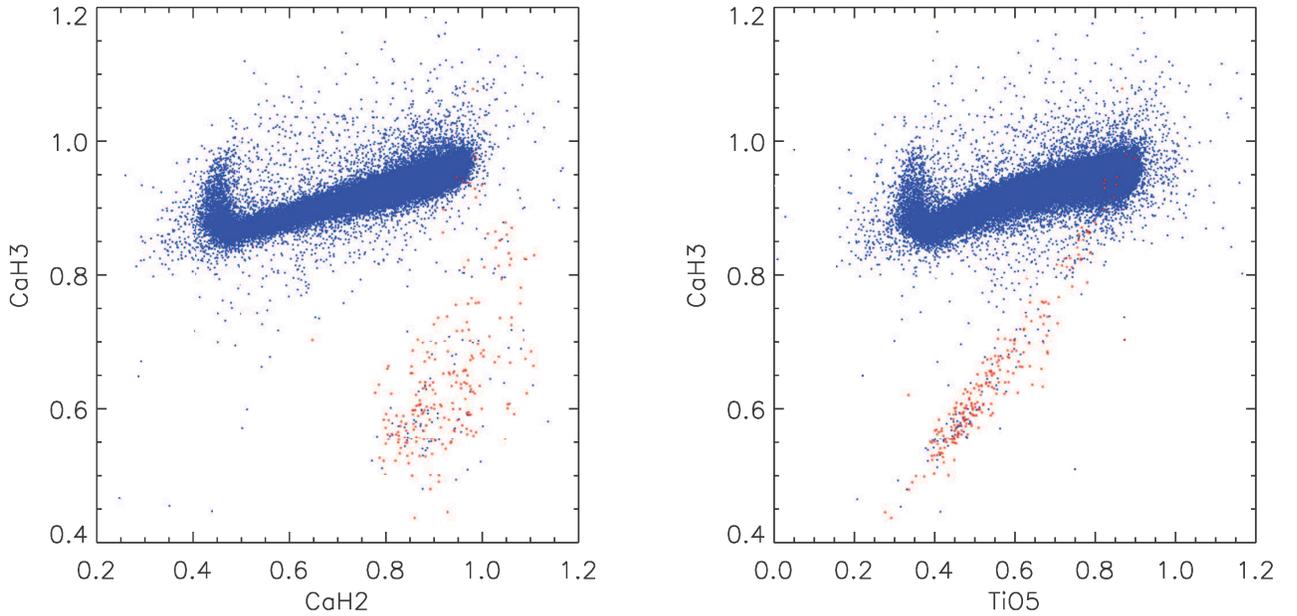}
   \caption{Spectral indices for {\it all} M giants. Blue and red dots represent M giants and confirmed carbon stars respectively. The concentrated distribution of carbon stars in the spectral indices diagram suggests that CaH indices are an efficient criteria for discriminating carbon stars in the M giants sample. }
   \label{idx}
\end{figure*}

\begin{figure*}
   \centering
   \includegraphics[angle=0,scale=0.4]{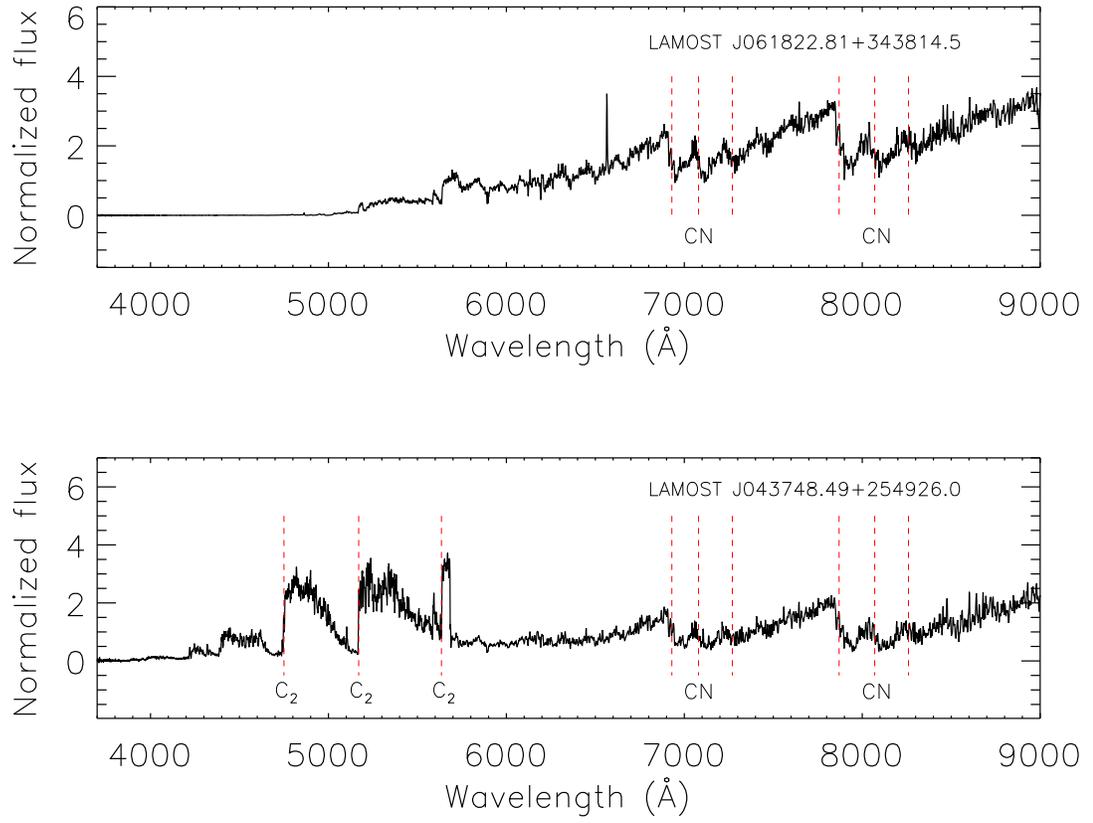}
   \caption{Example spectra of two representative kinds of carbon stars in our LAMOST derived catalog, which exhibit clear C$_2$ and CN molecular bands.}
   \label{cexp}
\end{figure*}

\begin{table*}
\caption{Description of the M giant and M dwarf catalog }.
\label{ctlg}
 \centering
\begin{tabular}{llcl}
\hline
Column &  Format  & Unit   & Description \\
\hline
DESIGNATION & string & - & object designation from the LAMOST DR5 \\
RA & float & deg & object right ascension in LAMOST DR5 (J2000)\\
DEC & float & deg & object declination in LAMOST DR5 (J2000)\\
CaH2 & float & -  & spectral index of LAMOST spectrum \\
CaH3 & float & -  & spectral index of LAMOST spectrum \\
TiO5 & float & -  & spectral index of LAMOST spectrum \\
SNR & float & - & signal to noise ratio of i filter in LAMOST spectrum \\
SpTy & string & -  & specral subtype classified by the template-fitting algorithm \\
Gmag$_0$ &  float & mag & dereddened G band magnitude from \gaia{} DR2\\
G$_{bp}$$_0$ &  float & mag & dereddened  BP band magnitude from \gaia{} DR2\\
G$_{rp}$$_0$ &  float & mag & dereddened  RP band magnitude from \gaia{} DR2\\
J$_0$ &  float & mag & dereddened J band magnitude from 2MASS\\
H$_0$ &  float & mag & dereddened H band magnitude from 2MASS\\
K$_{S0}$ &  float & mag & dereddened K$_S$ band magnitude from 2MASS\\
W1$_0$ &  float & mag & dereddened W1 band magnitude from \emph{WISE}\\
W2$_0$ &  float & mag & dereddened W2 band magnitude from \emph{WISE}\\
EBV & float & mag &  reddening from the 3D dust map\\
Dist & float & pc &  distance derived by the parallax in \gaia{} DR2 \\
RV & float & km s$^{-1}$ & radial velocity measured from LAMOST spectrum \\
$[M/H]$ & float &  - & estimated photometric metallicity of M giants\\
X & float & pc & Galactocentric coordinate points to the direction opposite to that of the Sun \\
Y & float & pc & Galactocentric coordinate points to the direction of Galactic rotation \\
Z & float & pc & Galactocentric coordinate points toward the North Galactic Pole   \\
U & float & km s$^{-1}$ & Galactic space velocity in X axis   \\
V & float & km s$^{-1}$ & Galactic space velocity in y axis    \\
W & float & km s$^{-1}$ & Galactic space velocity in Z axis   \\
SFLAG & string  & - & label of subsamples  \\
\hline
\end{tabular}
\end{table*}

\subsection{Carbon stars}
Carbon stars are peculiar objects which show an inversion of the C/O ratio (C/O $>$ 1). Since their spectra are dominated by strong carbon molecular bands (CH, CN or C$_2$), many of them are considered as asymptotic giant branch (AGB) stars undergoing the third dredge-up process. Moreover, in the revised Morgan–Keenan (MK) classification system \citep{1993PASP..105..905K}, nearly half of the carbon star sequence corresponds to ordinary oxygen late-type stars because of their low surface temperature. We believe that a small number of late-type carbon stars were included in the M-type catalog and mis-classified as M giants in our sample.

Using the LAMOST spectroscopic data, carbon stars have been systematically investigated by a few authors. \citet{2015RAA....15.1671S} found 183 carbon stars from the LAMOST pilot survey with an  efficient manifold ranking algorithm. In the LAMOST DR2 catalog, \citet{2016ApJS..226....1J} reported identifying 894 carbon stars, using a series of spectral indices (C$_2$ in 5,635 \AA, Ba II in 4,554 \AA, CN in 7,065 \AA ~  and 7,820 \AA) as the selection criteria. Using an efficient machine-learning algorithm, \citet{2018ApJS..234...31L} presented a catalog of 2,651 carbon stars from the LAMOST DR4. After de-duplicating these published catalogs, the number of carbon stars in LAMOST are 2,812 objects in total.

Combining with our M giants catalog and the published carbon stars catalog in LAMOST, 224 common stars were identified. Figure~\ref{idx} shows the spectral indices of our M giants catalog, where known carbon stars are marked as red dots. The concentrated distribution of carbon stars strongly suggests that the CaH spectral indices is an efficient criteria for selecting carbon stars in the M giants sample. To do this simply, we select carbon star candidates whose spectral indices satisfy the following criteria [CaH3 - 0.8 $\times $ CaH2 - 0.1] $<$ 0. Of the 314 carbon star candidates which passed through the selection criteria, 289 were confirmed as carbon stars by visual inspection of LAMOST spectra, while most of the excluded stars have low signal-to-noise ratio spectra. Two representative carbon spectra from LAMOST are shown in Figure~\ref{cexp} as examples.

\begin{table*}
\caption{Number of sources of the labeled subsample in our M-type catalog.}
\label{sub}
 \centering
\begin{tabular}{lcrr}
\hline
 Source type & ~Label &~~ M giants catalog  & ~ M dwarfs catalog\\
\hline
Confirmed M type stars   &   m  &  35,382  &  470,137 \\
Non-M type stars  & n & 1697 & 2224\\
Carbon stars & c & 289 & 0\\
White dwarf binaries & w & 15 & 57 \\
Unconfirmed M type stars &  u &  2413 & 28,734\\
\hline
Total & & 39,796 & 501,152 \\
\hline
\end{tabular}
\end{table*}

\section{Properties of M giants}
\subsection{Metallicity Estimation}
\begin{figure*}
   \centering
   \includegraphics[angle=0,scale=0.4]{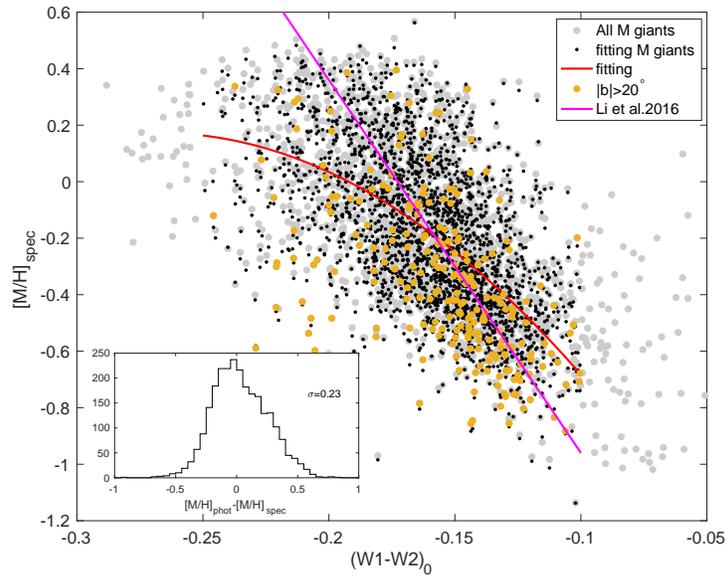}
   \caption{Metallicity distribution of APOGEE M giants vs. $(W1-W2)_0$ color. The red line shows the best-fit polynomial relationship. The inset histogram shows the scatter in metallicity about this relation, which has a dispersion of 0.23 dex.}
   \label{feh}
\end{figure*}

In this section, we compare the photometry of our M giants to the spectroscopic metallicity obtained using data from the SDSS project APOGEE \citep{holtzman2015}. This survey is taking high-S/N and high-resolution (R=22,500) NIR spectra, resulting in detailed chemistry and a measurement of [M/H] to a precision of better than 0.1 dex. As in \citet{2016ApJ...823...59L}, we cross-matched our M giant samples with APOGEE DR14, finding 2,895 stars in common, of which 2,689 are classified as true M giants, providing a much larger sample than used for fitting in \citet{2016ApJ...823...59L}.
Figure~\ref{feh} shows the APOGEE metallicities of the resulting sample of 2,689 M giants as a function of $(W1-W2)_0$ color. The correlation with $(W1-W2)_0$ can be fit with the following polynomial relation:

\begin{equation}
[M/H]=-30.57*(W1-W2)_0^2-16.34*(W1-W2)_0-2.01
\end{equation}

As can be seen from the inset of Figure~\ref{feh}, the residual scatter about this relation is 0.23 dex. We compare the new result to the previous similar work in \citet{2016ApJ...823...59L}, shown as a purple line in the figure.
It should be noted that our fits are only effective in the [M/H] range from $\sim -0.6$ dex to $\sim +0.2$ dex.

\begin{figure*}
   \centering
   \includegraphics[angle=0,scale=0.25]{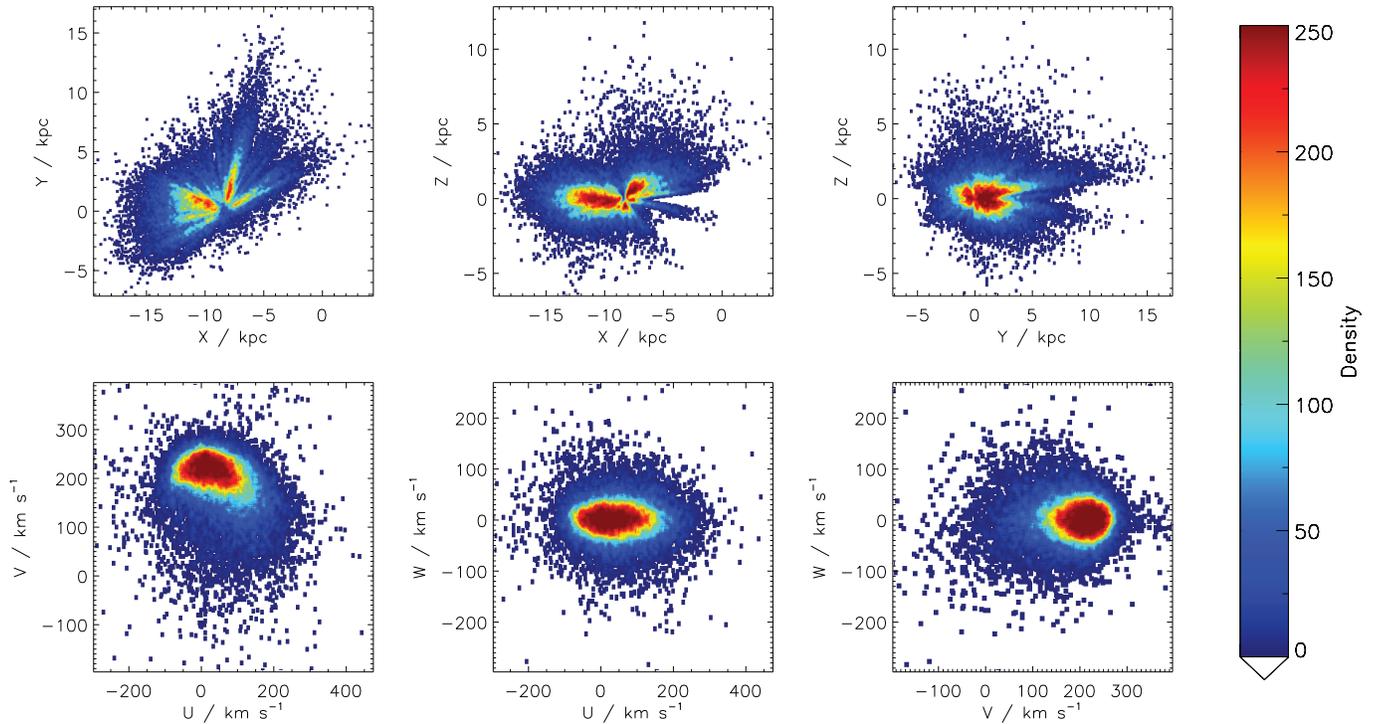}
   \caption{Galactocentric spatial distribution and space velocity distribution of confirmed M giants in LAMOST DR5. The spatial distribution clearly shows that the Galactocentric distances for most of M giants are less than 18 kpc. }
   \label{map}
\end{figure*}

\begin{figure*}
   \centering
   \includegraphics[angle=0,scale=0.25]{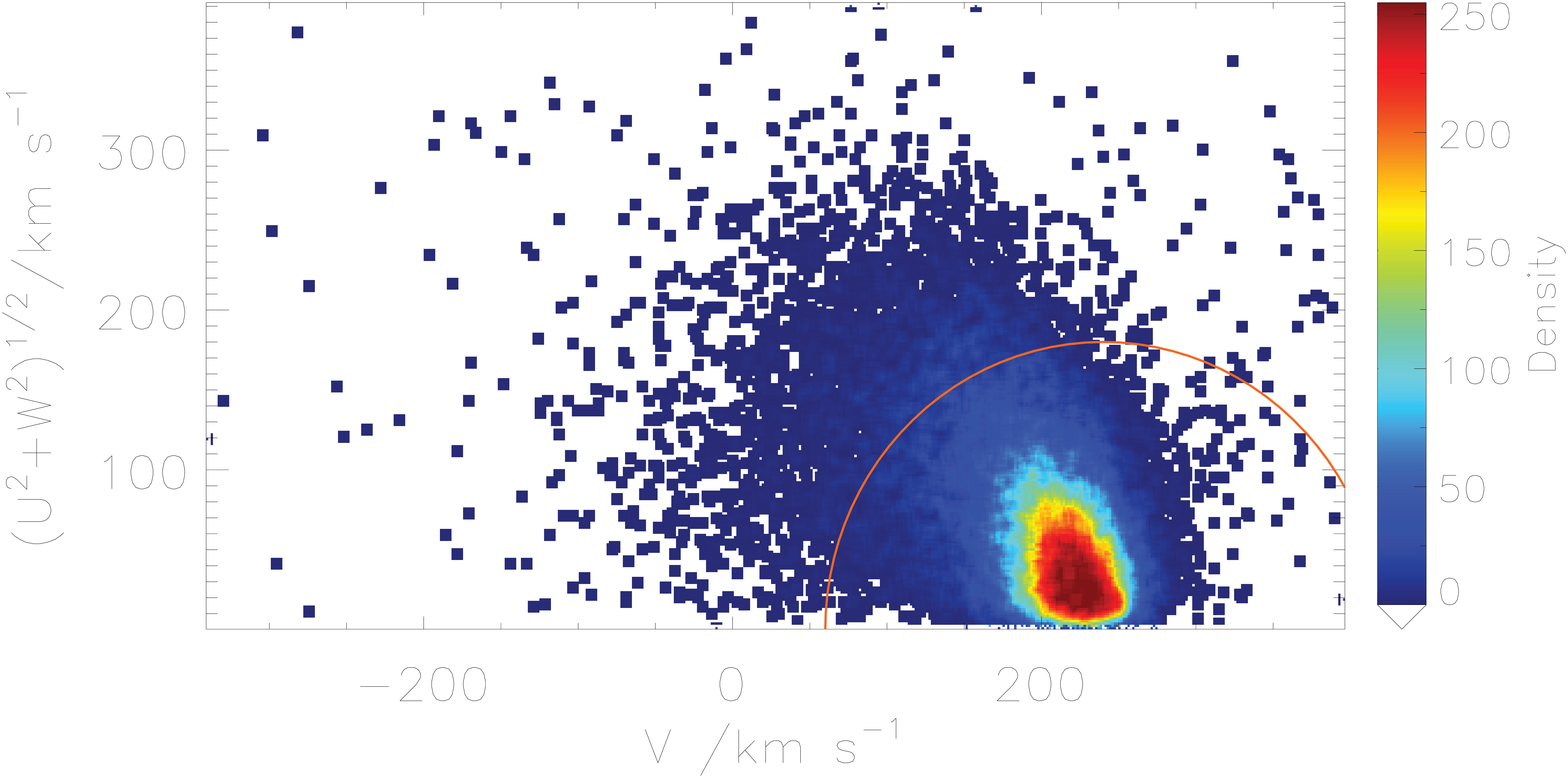}
   \caption{Toomre density diagram of confirmed M giants. The red line kinematically divides the M giants sample into halo and disk components, since stars with $\mid$V - V$_{LSR} \mid  >$ 180 km s$^{-1}$ have higher probabilities of being members of the Galactic halo. }
   \label{toomre}
\end{figure*}


\subsection{Galactocentric coordinates and space motions}

We calculate the Galactocentric coordinates (X,Y,Z) and space velocities (U,V,W) for all confirmed M giants. The solar position and the circular velocity at the solar location are adopted as R$_0$=-8.34 kpc and V$_c$=240 km~s$^{-1}$ \citep{2014ApJ...783..130R}. To correct for the solar motion, we adopt the peculiar velocity of the Sun in the local standard of rest (U$_\odot$, V$_\odot$, W$_\odot$)=(11.1, 12.24, 7.25) km s$^{-1}$ \citep{2010MNRAS.403.1829S}.

Figure~\ref{map} shows the Galactocentric spatial distribution and space velocity distribution for confirmed M giants. Since a large part of LAMOST observations are focused on the Galactic Anti-center, most of the identified M giants in our catalog are also located on this region. Another prominent clump of M giants is in the $Kepler$ field, centered on $l=76.38^{\circ}$, $b=10.8^{\circ}$, which has been observed repeatedly by LAMOST. The spatial distribution of M giants clearly shows that most of M giants are located on the Galactic disk and that the Galactocentric distances are less than 18 Kpc.

We summarize the kinematic components of the M giants sample with a Toomre diagram in Figure~\ref{toomre}. The halo stars are defined as having $\mid$V - V$_{LSR} \mid  >$ 180 km s$^{-1}$ \citep{2004AJ....128.1177V, 2009MNRAS.399.1145S}, where V$_{LSR}$=(0,240,0) km s$^{-1}$. The total number of halo stars selected in this manner is 5,334, constituting about 14$\%$ of our M giants sample.


\section{Catalog description}\label{catalog}
The complete catalog of M-type stars is provided in the electronic version of this paper, including 39,796 M giants and 501,152 M dwarfs. As described in \citet{2015AJ....150...42Z}, the classification results contain M giants with 7 temperature subtypes from M0 to M6, as well as M dwarfs with 18 temperature subtypes from K7.0 to M8.5 and 12 metallicity subclasses from dMr to usdMp. Since the spectral templates were corrected to the rest frame, the radial velocity of M giants and M dwarfs were calculated with a measurement uncertainty of about 5 km s$^{-1}$ \citep{2015RAA....15.1154Z}. Spectral indices of CaH2, CaH3 and TiO5 were provided for measuring molecular features, which were defined by \citet{1995AJ....110.1838R} and \citet{2007ApJ...669.1235L}. In addition, our catalog includes photometry in optical bands Gmag/G$_{bp}$/G$_{rp}$ from \gaia{} DR2, near infrared bands J/H/K$_S$ from 2MASS, infrared bands W1/W2 from \emph{WISE}, and distance as derived from the parallax in \gaia{} DR2 \citep{2018AJ....156...58B}. In Section~\ref{mstar}, the de-reddened magnitudes $Gmag_0$/$G_{bp0}$/$G_{rp0}$/$J_0$/$H_0$/$Ks_0$/$W1_0$/$W2_0$ and absolute magnitude $M_G$ for each star were calculated, thus all these derived magnitudes are provided in our catalog. In Table~\ref{ctlg} we provide the description of each column of our catalog.

\begin{figure*}
   \centering
   \includegraphics[angle=0,scale=0.4]{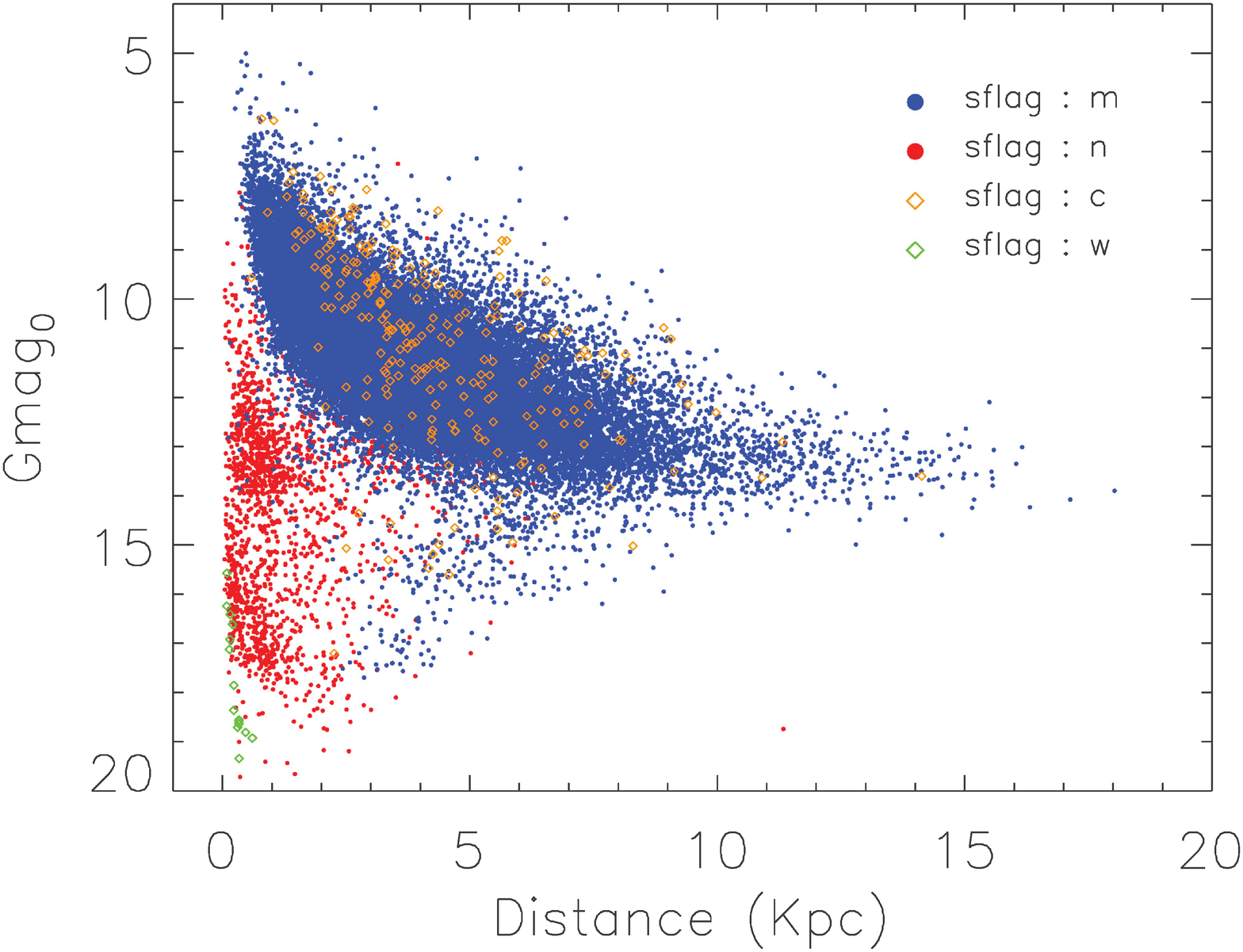}
   \caption{De-reddened Gmag Vs. distance diagram for all M giants in our catalog. Colors represent different subsamples which were discussed in Section~\ref{results}. For most of the confirmed M giants, the distance is between 1 to 15 kpc, with a peak at 2.5 kpc. Taking into account the intrinsic luminosity of different subsamples, the distance distribution of each subsample is consistent with the expectation.}
   \label{dist}
\end{figure*}

As discussed in Section~\ref{results}, there are non-M type stars both in the M giants and M dwarfs catalog. In order to mark those non-M type stars and further purify the M type stars sample, we have added a column called 'sflag' in the M-type catalog, including 'm' as confirmed M type stars, 'n' as non-M type stars, 'c' as carbon stars, 'w' as white dwarf binaries, and 'u' as unconfirmed M type stars because of the lack of \gaia{} data. Table~\ref{sub} shows the number of sources in each labeled subsample.

\section{summary}
 We have used the template-fitting pipeline of our previous works to identify and classify M dwarfs and M giants in LAMOST DR5. A total of 39,796 M giants and 501,152 M dwarfs are provided. In addition to the large number of newly classified M-type stars, we demonstrate that M giants and M dwarfs can be well separated in the spectral indices diagram as well as in the infrared color-color diagram. However, using the \gaia{} DR2 data, the M$_G$ vs. (G$_{bp}$-G$_{rp}$)$_0$ diagram clearly shows that not only M  type stars, but also early-type stars, and white dwarf-M dwarfs binaries were included in the M type sample. The contamination rates in the M giants and M dwarfs samples are about 4.6\% and 0.48\% respectively.  In the published M-type catalog, 'sflag' is used to mark the confirmed M type stars as well as other contamination sources. In particular, we found that the CaH spectral indices are efficient selection criteria for carbon stars. Finally, about 289 carbon stars were selected from the M giants sample, and further confirmed through visual inspection of LAMOST spectra.

To verify the reliability of our classification results in the M giants catalog, we plot the dereddened Gmag vs. distance diagram in Figure~\ref{dist}. Colors represent different subsamples labeled in Table~\ref{sub}. The distance distribution of M giants extends from 1 to 15 kpc, with a peak at 2.5 kpc. As expected, faint stars with small distances were labeled as non-M giants in our catalog, including early-type stars, M dwarfs, and white dwarf binaries. A few M giants are located in the outlier region with Gmag$_0$ greater than 14 and distance from 3 to 8 kpc. This is because the faint magnitude makes the parallax uncertainty of these M giants relatively large, which further leads to unreliable distance estimation. In addition, spectral visual inspection confirms that these outliers are M giants.

\acknowledgments
This work is supported by National Natural Science Foundation of China (NSFC) under grants U1731129 and 11503066 (PI: J.Zhong). JL would like to acknowledge the NSFC under grant 11703019 and China West Normal University grants 17C053,17YC507 and 16E018. JLC acknowledges support from the U.S. National Science Foundation via grant AST-1816196. RAM acknowledges support from the Chilean Centro de Excelencia en
Astrofisica y Tecnologias Afines (CATA) BASAL AFB-170002, and FONDECYT/CONICYT grant \# 1190038.

Guoshoujing Telescope (the Large Sky Area Multi-Object Fiber Spectroscopic Telescope LAMOST) is a National Major Scientific Project built by the Chinese Academy of Sciences. Funding for the project has been provided by the National Development and Reform Commission. LAMOST is operated and managed by the National Astronomical Observatories, Chinese Academy of Sciences.

This work is sponsored (in part) by the Chinese Academy of Sciences (CAS), through a grant to the CAS South America Center for Astronomy (CASSACA) in Santiago, Chile.

This work has made use of data from the European Space Agency (ESA)mission {\it Gaia} (\url{https://www.cosmos.esa.int/gaia}), processed by the {\it Gaia} Data Processing and Analysis Consortium (DPAC,\url{https://www.cosmos.esa.int/web/gaia/dpac/consortium}). Funding for the DPAC has been provided by national institutions, in particular the institutions participating in the {\it Gaia} Multilateral Agreement.





\begin{thebibliography}{}


\bibitem[Bailer-Jones et al.(2018)]{2018AJ....156...58B} Bailer-Jones, C.~A.~L., Rybizki, J., Fouesneau, M., Mantelet, G., \& Andrae, R.\ 2018, \aj, 156, 58

\bibitem[Bochanski et al.(2014)]{2014AJ....147...76B} Bochanski, J.~J., Willman, B., West, A.~A., Strader, J., \& Chomiuk, L.\ 2014, \aj, 147, 76

\bibitem[Casagrande \& VandenBerg(2018)]{2018MNRAS.479L.102C} Casagrande, L., \& VandenBerg, D.~A.\ 2018, \mnras, 479, L102

\bibitem[Chen et al.(2001)]{2001ApJ...553..184C} Chen, B., Stoughton, C., Smith, J.~A., et al.\ 2001, \apj, 553, 184.

\bibitem[Chou et al.(2007)]{2007ApJ...670..346C} Chou, M.-Y., Majewski, S.~R., Cunha, K., et al.\ 2007, \apj, 670, 346

\bibitem[Cui et al. (2012)]{Cui2012} Cui, X.-Q., Zhao, Y.-H., Chu, Y.-Q., et al., RAA, 12, 1197

\bibitem[Gaia Collaboration et al.(2018)]{2018A&A...616A...1G} Gaia Collaboration, Brown, A.~G.~A., Vallenari, A., et al.\ 2018, \aap, 616, A1.

\bibitem[Gaia Collaboration et al.(2018)]{2018A&A...616A..10G} Gaia Collaboration, Babusiaux, C., van Leeuwen, F., et al.\ 2018, \aap, 616, A10

\bibitem[Green et al.(2018)]{2018MNRAS.478..651G} Green, G.~M., Schlafly, E.~F., Finkbeiner, D., et al.\ 2018, \mnras, 478, 651

\bibitem[Holtzman et al.(2015)]{holtzman2015} Holtzman, J.~A., Shetrone, M., Johnson, J.~A., et al.\ 2015, \aj, 150, 148

\bibitem[Ji et al.(2016)]{2016ApJS..226....1J} Ji, W., Cui, W., Liu, C., et al.\ 2016, \apjs, 226, 1

\bibitem[Keenan(1993)]{1993PASP..105..905K} Keenan, P.~C.\ 1993, \pasp, 105, 905

\bibitem[Koposov et al.(2014)]{Koposov2014} Koposov, S.~E., Belokurov,V., Zucker, D.~B., et al.\ 2014, MNRAS, 446, 3110

\bibitem[L{\'e}pine et al.(2007)]{2007ApJ...669.1235L} L{\'e}pine, S., Rich, R.~M., \& Shara, M.~M.\ 2007, \apj, 669, 1235

\bibitem[Li et al.(2016a)]{2016ApJ...823...59L} Li, J., Smith, M.~C., Zhong, J., et al.\ 2016, \apj, 823, 59

\bibitem[Li et al.(2016b)]{2016RAA....16..125L} Li, J., Liu, C., Carlin, J.~L., et al.\ 2016, Research in Astronomy and Astrophysics, 16, 125


\bibitem[Li et al.(2018)]{2018ApJS..234...31L} Li, Y.-B., Luo, A.-L., Du, C.-D., et al.\ 2018, \apjs, 234, 31

\bibitem[Li et al.(2019)]{2019arXiv190207861L} Li, J., Liu, C., Xue, X., et al.\ 2019, arXiv:1902.07861

\bibitem[Luo et al.(2012)]{Luo2012} Luo, A.-L., Zhang, H.-T., Zhao, Y.-H., et al.\ 2012, Research in Astronomy and Astrophysics, 12, 1243
\bibitem[Majewski et al.(2003)]{2003ApJ...599.1082M} Majewski, S.~R.,Skrutskie, M.~F., Weinberg, M.~D.,\& Ostheimer, J.~C.\ 2003, \apj, 599, 1082

\bibitem[Minniti et al.(2011)]{2011ApJ...733L..43M} Minniti, D., Saito, R.~K., Alonso-Garc{\'\i}a, J., et al.\ 2011, \apj, 733, L43.

\bibitem[Reid et al.(1995)]{1995AJ....110.1838R} Reid, I.~N., Hawley, S.~L., \& Gizis, J.~E.\ 1995, \aj, 110, 1838

\bibitem[Reid et al.(2014)]{2014ApJ...783..130R} Reid, M.~J., Menten, K.~M., Brunthaler, A., et al.\ 2014, \apj, 783, 130.

\bibitem[Robin et al.(1992)]{1992ApJ...400L..25R} Robin, A.~C., Creze, M., \& Mohan, V.\ 1992, \apj, 400, L25.

\bibitem[Sch{\"o}nrich, \& Binney(2009)]{2009MNRAS.399.1145S} Sch{\"o}nrich, R., \& Binney, J.\ 2009, \mnras, 399, 1145.

\bibitem[Sch{\"o}nrich et al.(2010)]{2010MNRAS.403.1829S} Sch{\"o}nrich, R., Binney, J., \& Dehnen, W.\ 2010, \mnras, 403, 1829.

\bibitem[Si et al.(2015)]{2015RAA....15.1671S} Si, J.-M., Li, Y.-B., Luo, A.-L., et al.\ 2015, Research in Astronomy and Astrophysics, 15, 1671

\bibitem[Skrutskie et al.(2006)]{2006AJ....131.1163S} Skrutskie, M.~F., Cutri, R.~M., Stiening, R., et al.\ 2006, \aj, 131, 1163


\bibitem[Venn et al.(2004)]{2004AJ....128.1177V} Venn, K.~A., Irwin, M., Shetrone, M.~D., et al.\ 2004, \aj, 128, 1177.

\bibitem[Wright et al.(2010)]{2010AJ....140.1868W} Wright, E.~L., Eisenhardt, P.~R.~M., Mainzer, A.~K., et al.\ 2010, \aj, 140, 1868

\bibitem[Yuan et al.(2013)]{2013MNRAS.430.2188Y} Yuan, H.~B., Liu, X.~W., \& Xiang, M.~S.\ 2013, \mnras, 430, 2188
\bibitem[Zhao et al.(2012)]{Zhao2012} Zhao, G., Zhao, Y.-H., Chu,
  Y.-Q., et al. \ 2012, RAA, 12, 723
\bibitem[Zhong et al.(2015a)]{2015AJ....150...42Z} Zhong, J., L{\'e}pine, S., Hou, J., et al.\ 2015, \aj, 150, 42

\bibitem[Zhong et al.(2015b)]{2015RAA....15.1154Z} Zhong, J., L{\'e}pine, S., Li, J., et al.\ 2015, Research in Astronomy and Astrophysics, 15, 1154

\end{thebibliography}
\end{document}